\documentclass[%
 aip,
 amsmath,amssymb,
 reprint,%
]{revtex4-1}
\usepackage{graphicx}
\usepackage{dcolumn}
\usepackage{bm}

\usepackage[utf8]{inputenc}
\usepackage[T1]{fontenc}
\usepackage{mathptmx}
\usepackage{etoolbox}
\usepackage{natbib}
\makeatletter
\def\@email#1#2{%
 \endgroup
 \patchcmd{\titleblock@produce}
  {\frontmatter@RRAPformat}
  {\frontmatter@RRAPformat{\produce@RRAP{*#1\href{mailto:#2}{#2}}}\frontmatter@RRAPformat}
  {}{}
}%
\makeatother
\begin{document}

\preprint{AIP/123-QED}

\title[High viscosity effects on solitary wave propagation]{High viscosity effects on solitary wave propagation}
\author{C. Calvo}
\author{A. Tamburrino}%
 \altaffiliation[Also at ]{Advanced Mining Technology Center, AMTC, Universidad de Chile, Santiago, Chile.}
\affiliation{ 
Departamento de Ingeniería Civil, Facultad de Ciencias Físicas y Matemáticas, Universidad de Chile, 
Blanco Encalada 2002, 8370449 Santiago, Chile}%
\author{C.Falc\'on}
\email{cfalcon@uchile.cl}
\affiliation{%
Departamento de Física, Facultad de Ciencias Físicas y Matemáticas, Universidad de Chile, Blanco Encalada 2008, 8370449 Santiago, Chile
}%

\date{\today}

\begin{abstract}
We present an experimental and numerical study of linear and non-linear viscous effects in transient non-linear long wave propagation in Newtonian and shear thinning fluids in the laminar flow regime. Using optical measuring techniques (Fourier Transform Profilometry) and numerical simulations (open- source CFD library OpenFOAM), we show that the wave phase speed decreases in both glycerin and carboxymethylcellulose (CMC) solutions with respect to that in water. A decrease in wave phase speed is observed, and a dispersion relation is obtained for surface waves through dimensional analysis from five dimensionless groups: the dimensionless wave celerity, the shallowness parameter, dimensionless amplitude, Reynolds number and the flow index. To complete the picture on wave propagation, an empirical dependence between the wave attenuation and the last four dimensionless groups mentioned above is found for non-linear long surface waves. We conclude quantitatively about all the viscosity effects in non-linear long wave propagation.
\end{abstract}

\maketitle

%

\section{Introduction}\label{sec:lntro}

Solitary waves, i.e. transient non-linear long waves, were first observed by John Scott Russell propagating over the surface of water while ”preserving its original figure”~\cite{solitonRussell} and are sustained by the balance between nonlinear advection and frequency dispersion. Solitary wave propagation was later found to be ubiquitous in many different contexts, ranging from the propagation of solitary pulses in fiber optics~\cite{hasegawa1973} to solitary wave propagation of spin textures in magnetic ribbons~\cite{kosevitch1990}, just to mention a few examples~\cite{remoissenetBook,belashov2006}. In the theoretical description of solitary wave propagation, the effect of dissipation is usually treated as a perturbation. This approach is troublesome when dealing with very viscous Newtonian fluids, and in the case of non-Newtonian ones the nonlinear nature of the viscous dissipation prevents it completely. Even so, solitary waves are observed propagating at the surface of viscous Newtonian fluids when viscosity cannot be neglected~\cite{kakutani1975,liu2004,liu2006} In the case of non-Newtonian fluids, solitary waves have been predicted and numerically simulated on melted rock inside the crust of the earth and in some geophysical processes involving fluid devolatilization~\cite{gleesonBook,connolly2015}. The rheology of these geophysical fluids supporting these waves is expected to be nonlinear according to Ref.~\cite{ranalli1995}. Some other cases of waves in different rheological media comprise solitary waves in pulsating flow inside blood vessels~\cite{yomosa1987} and viscous dissipation of atmospheric solitons in the earth ionosphere caused by earth seismic surface waves~\cite{belashov2006}.

Solitary waves propagating over the surface of a fluid have a particular $sech^{2}$-shape. This shape arises as a nonlinear solution from the averaged Euler equations, either obtained as an exact solution of the Korteweg-de Vries equations or the Boussinesq system~\cite{johnson1997}. Weakly linear viscosity effects had been taken into account for finding wave solutions from the Navier-Stokes equations. Indeed, novel theoretical, experimental and numerical studies addressed the weak nonlinearity tied to the Boussinesq assumptions~\cite{liu2004,liu2006,johnson1997,mei1998,behroozi2003,winckler2015,klettner2012}. All of them have used the boundary layer theory in order to decompose the velocity field into a rotational and a potential field. Therefore, the Boussinesq solution for the solitary wave holds along the irrotational zone between two boundary layers while the unsteady boundary layer equation is solved in the rotational zones, thus obtaining the weakly viscous wave damping. These assumptions do not hold for highly viscous or non-Newtonian fluids where the flow is not irrotational in most of the flow domain. Thus, quantifying and explaining the viscous dissipation in long waves remain a challenge.

To our knowledge, linear and non-linear highly viscous effects have not been fully addressed and understood on these waves,  due to the complexity of the Navier-Stokes equations, even in the laminar regime. Semi analytical and numerical studies have been performed so as to explain the propagation of these waves in nature \citep{connolly2015}. However, a complete Navier-Stokes formulation and experimental studies are needed in order to explain quantitatively and qualitatively the fluid dynamics of this phenomena.   

The purpose of the present experimental study is to improve the understanding of the role of different viscous effects on non-linear transient long wave propagation, quantifying and relating wave parameters with rheology variables. The main wave parameter under study is the speed of the maximum height of the nonlinear wave, which we will call wave speed $C$. Although this definition is not rigorous as wave speed is defined for linearized equations, we will use it as a way to characterize the wave propagation. To quantify the relation between $C$ and the rheological variables of the viscous fluids, we based our study on experimental measurements of propagative solitary waves in Newtonian and non-Newtonian fluids using non intrusive optical techniques and corroborated these findings with computational fluid dynamics simulations in order to quantify the highly viscous linear and nonlinear effects on wave phase speed and damping. This is done in order to complement the existing theory for weakly non-linear viscous effects on these waves.

This article is organized as follows. The fundamental governing equations are presented in Section~\ref{sec:gob_eqs}, the experimental and numerical procedures are explained in Section~\ref{sec:exp} and \ref{sec:num}, respectively. These results are presented in Section~\ref{sec:res} where we show the obtained linear and non-linear viscous effects on non-linear wave propagation. Finally in Section~\ref{sec:dis} we discuss these results and demonstrate that linear viscosity tends to reduce the wave phase speed whereas a lower non-linearity index contributes to increase this velocity in non-Newtonian fluids. 

\section{Governing equations}\label{sec:gob_eqs}

\subsection{Fluid dynamic equations}

\begin{figure}
\centering
 \includegraphics[width=1.0\columnwidth]{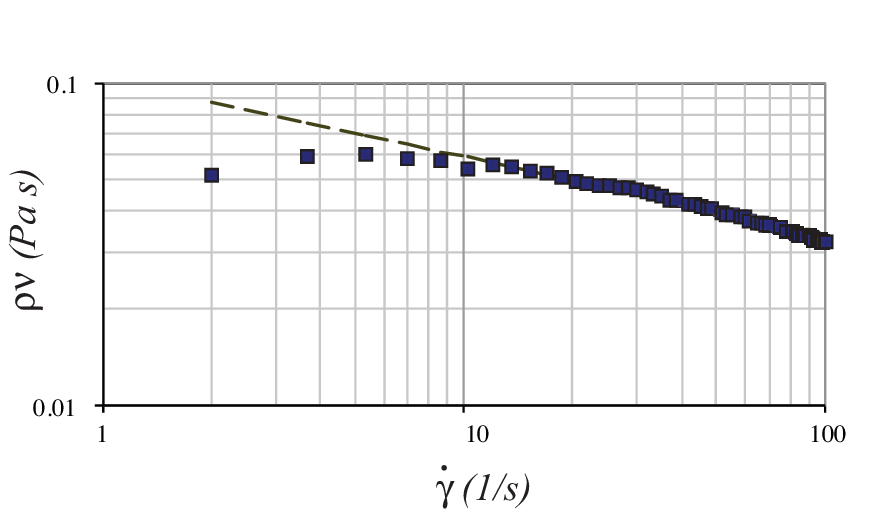}
\caption{ Experimental measurement of the shear viscosity $\rho \nu$ as a function of the shear rate $\dot{\gamma}$ for the CMC solution used in our experiments. Here the dashed line is a least square fit of a power-law behavior between 10 to 100 s$^{-1}$ with slope $n-1$=-0.24$\pm$ 0.02 and $K$=0.099$\pm$0.004. Errorbars are omitted as they lie within the symbols. Only increasing ramps in shear rate are shown. }
\label{fig:shear}
\end{figure}

Mass conservation and momentum balance for the incompressible Navier-Stokes equations are used to describe the propagation of  non-linear waves with wavelenth $\lambda$. These equations read 

\begin{equation}
\nabla \cdot \mathbf{u}=0
\label{eq:continuity}
\end{equation}

\begin{equation}
\frac{\partial {\mathbf{u}}}{\partial t}+\left( {\mathbf{u}} \cdot \nabla \right) {\mathbf{u}} =-\frac{1}{\rho} \mathbf{\nabla} \left( p+\rho g h \right)+\frac{1}{\rho} \mathbf{\nabla} \cdot f({\mathbf{u}})
\label{eq:momentum}
\end{equation}
where $\mathbf{u}$ is the velocity field, $p$ the pressure $\rho$ the mass density, $g$ is gravity and $h$ is the local depth of the fluid system. Surface tension effects quantified by the interfacial tension $\sigma$ are neglected as the Bond number $Bo = (\rho g\lambda^{2} /\sigma)^{1/2} \ll 1$ for the configurations used in this work. The stress tensor in equation \ref{eq:momentum} can be modeled in several ways depending on their observed rheology~\cite{larsonBook}. From our experimental data presented in Fig. 1, in the range $\dot{\gamma}\in[10 ,100]$ s$^{-1}$, the stress tensor can be modeled by a power-law model as

\begin{equation}
f({\mathbf{u}})=K \left( \frac{\left( \nabla \mathbf{u}+{\nabla \mathbf{u}}^T \right) : \left( \nabla \mathbf{u}+{\nabla \mathbf{u}}^T \right)
}{2} \right)^{\frac{n-1}{2}}  \left( \nabla \mathbf{u}
+{\nabla \mathbf{u}}^T \right)
\label{eq:nn}
\end{equation}
where $K$ and $n$ are the rheology consistency and flow index of a non-Newtonian fluid, respectively. Here $K$ has units of $[Pas^n]$ and $n$ dimensionless, as described above. In the case where $K=\rho\nu$ and $n=1$, one recovers the stress tensor for a Newtonian fluid. In Fig.~\ref{fig:shear} we show the typical shear viscosity measurements that give $n$ and $K$ for the shear thinning fluid used in our experiments.  It is observed that the power law model fits well for shear rates  $\dot{\gamma}$ greater than 10 s$^{-1}$, that was considered valid in the complete range of  $\dot{\gamma}$ for simplicity in the analysis.

\subsection{Non-dimensionalization}\label{sec:dim_a}

Defining $\varepsilon=H/h\ll1$ as the ratio of wave height to still fluid depth, we normalize the horizontal velocity and vertical velocity as $\tilde{u}=u /\varepsilon \sqrt{gh}$ and $\tilde{w}=w kh / \varepsilon \sqrt{gh}$, respectively. Here we have used $k=2\pi/\lambda$ as a wave number with $\lambda$ the typical width of the nonlinear wave. In our experiments, $\lambda$ is of the order of 3-5 cm depending on the wave height. With this definition, a shallowness parameter can be defined as $\mu=k h\ll1$, which quantifies the long wavelength approximation used for shallow water wave equations~\citep{liu2004}. The $x$ coordinate along the solitary wave propagation direction is then scaled to $\tilde{x}=kx=\mu x/h$, and the transverse coordinate is scale accordingly $\tilde{y}=\mu y/h$. The free surface deformation $\eta$ is scaled by the solitary wave height $\tilde{\eta}=\eta /H$, while time is scaled by the typical lineal frequency of the wave as $\tilde{t}=t k \sqrt{gh}$. To note all these variables we use $\tilde{(\cdot)}$ which denotes rescalling. Substituting the rescaled variables in equation \ref{eq:momentum}, averaging over the layer depth and following the prescription of Liu~\cite{liu2004,liu2006} to take into account the boundary layer separation, the vertical coordinate needs to be rescaled as $\tilde{z}=\mu z  / (h\sqrt{\bar{Re}})$ to preserve the order of magnitude of the dissipative terms in Eq. (\ref{eq:momentum}). Thus, in this configuration a rescaled Korteweg-de Vries equation with a source term stemming from the boundary layer contribution is found\cite{liu2004}. In this scaling, we use $\bar{Re}=\sqrt{gh} h /\nu$ as the modified Reynolds number for Newtonian fluids~\cite{liu2004,johnson1997}.
In the case of shear thinning fluids (and non-Newtonian fluids in general), we extend the same procedure used by Liu~\cite{liu2006} using a modified Reynolds number 
\begin{equation}
\widehat{\bar{Re}}=\left(\rho \sqrt{gh} h^n K^{-1} (\varepsilon \sqrt{gh})^{1-n} \right)^{\frac{2}{n+1}}\mu^{\frac{1-n}{n+1}}
\end{equation}
 to allow the same order of magnitude of the viscous dissipative nonlinear terms. It must be noticed that $\widehat{\bar{Re}}=\bar{Re}$ for $K=\rho\nu$ and $n=1$. From the Korteweg-de Vries solitary wave equation in \cite{liu2004}, the wave speed can be found by linearizing its solution in the weakly visocus limit and, in accordance to the dispersion relation for linear waves in viscous fluids, the wave phase speed must decrease with the square root of the inverse of the modified Reynolds number \citep{johnson1997}. Thus, the dimensionless wave speed ($\tilde{C}=c/\sqrt{gh}$) for the nonlinear wave propagating in the limit of small $\varepsilon$ in a non-Newtonian fluid is 

\begin{equation}
\widetilde{C}=1+\varepsilon/2- \frac{\chi}{(\mu \widehat{\bar{Re}})^{1/2}}
\label{eq:dispersion}
\end{equation}
where $\chi$ is a fitting constant obtained experimentally from a non-linear least squares regression, as described below. In the case of a Newtonian fluid it suffices to take $n=1$ and $K=\rho\nu$ \citep{liu2004}. The possible functional dependence of $\chi$ on $\varepsilon, \mu, \widehat{\bar{Re}}$ and $n$ might be computed from a nonlinear analysis similar to the followed here,  but it is outside of the scope of this work.

\section{Methods}\label{sec:exp}

\subsection{Experimental procedure}

\begin{figure*}
\centering
\begin{minipage}[c]{0.6\linewidth}
  \centering
  \includegraphics[width=\textwidth]{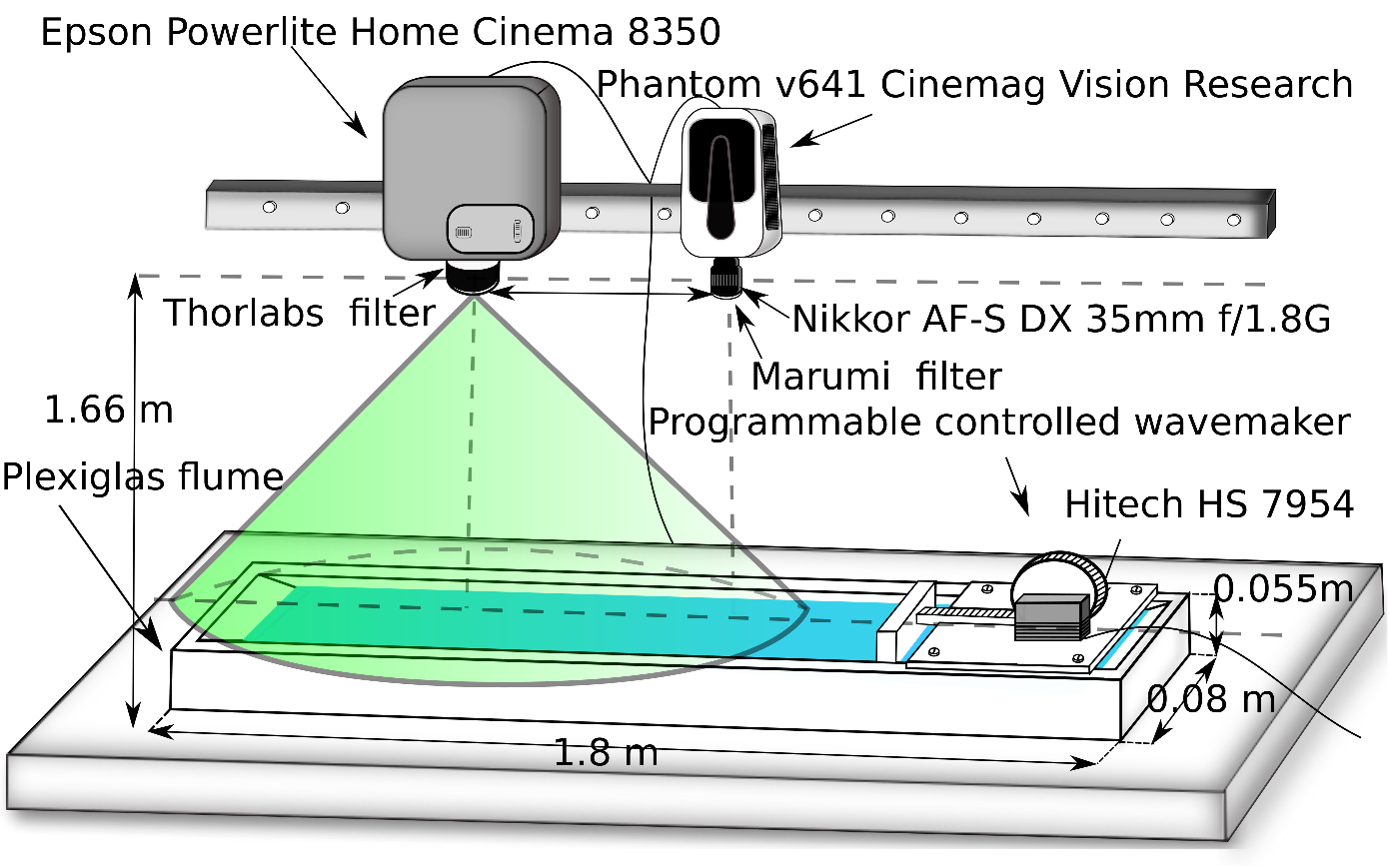}
\end{minipage}%
\begin{minipage}[c]{0.4\linewidth}
  \centering
  \includegraphics[width=\textwidth]{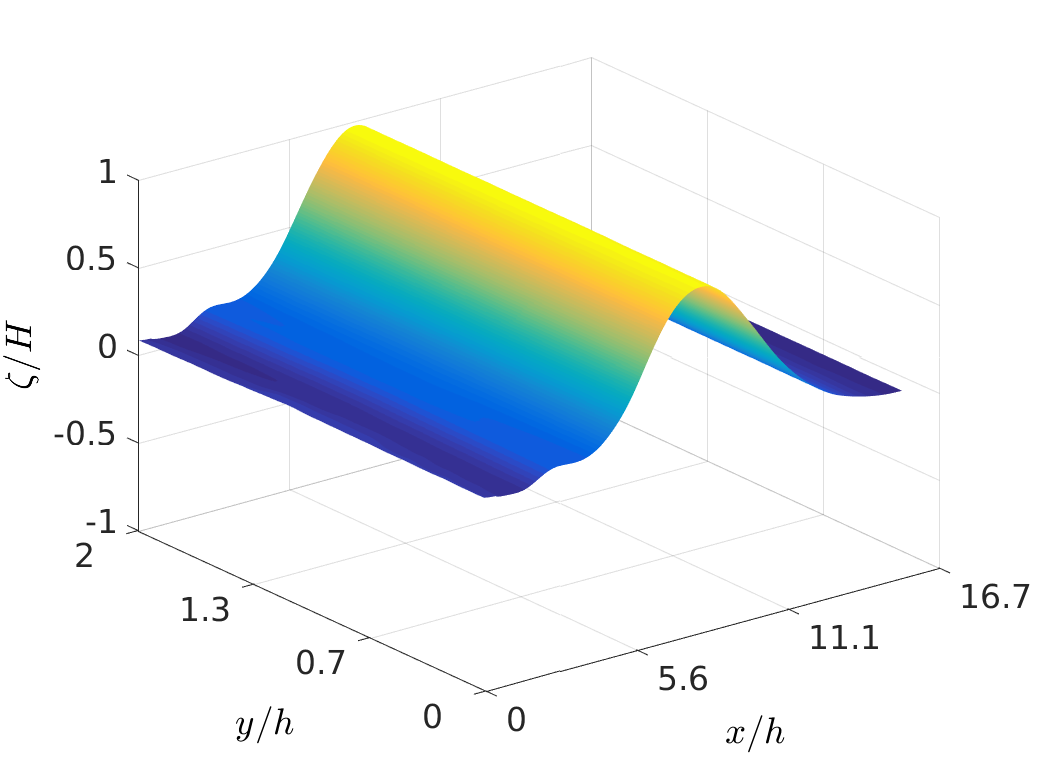}
\end{minipage}
\caption{{\bf Left:} Experimental setup. {\bf Right:}  Measured solitary wave. The reference $x=0$ is the resting paddle position. We rescale the longitudinal and transverse direction as $x/h$ and $y/h$, respectively.}
\label{fig:setup}
\end{figure*}

The experimental scheme is displayed in Fig.~\ref{fig:setup}(left). Experiments were conducted in a 1.800 m long, 0.080 m wide and 0.055 m tall rectangular Plexiglas flume. Four different still fluid depths were used: 0.020 m, 0.025 m, 0.030 m, 0.035 m and 0.040 m. Three different wave heights were measured at each water depth. Wave height to water depth ratios ranged from 0.05 to 0.49. Three different viscous fluids were used; distilled water, 71$\%$ aqueous USP, 99.5$\%$ glycerin solution and a 0.2$\%$ carboxymetilcellulose (CMC) solution. Their physical properties are presented in table~\ref{TableI}. The density $\rho$, and the rheological properties of the fluids were measured in the laboratory. For glycerin and the CMC solution $\rho$ was measured using a densimeter and $K$ and $n$ were measured using a temperature controlled Anton Paar Rheolab QC rheometer with a double gap measurement system. In the case of water, density and viscosity were calculated from the work of Keslin~\cite{keslin1978}. Waves were generated by a programmable controllable piston-type wavemaker completely inmersed within the fluid layer, which performed a horizontal motion with a $sech^{2}$-shape. The wave maker was driven by a digital servo motor (Hitec HS-7954HS, 7.4 V) attached to a gear mechanism, transforming circular to vertical motion. The digital motor was controlled using a microcontroller (Arduino Uno in a Teensy 3.1 USB board). The paddle motion was programmed following experimental procedures already developed to geenerate contolled wave profiles~\cite{malek2010,wu2014}, allowing the generation of a controlled impulse response of the water level. In order to measure the free surface displacement, an optical 2D non intrusive technique known as Fourier Transform Profilometry~\cite{takeda1982,maurel2009} was used. Experiments were processed from captured frames filmed with a high speed camera (Phantom v641 Cinemag Vision Research) recording at 300 fps, setting a 1616x330 pix window with a resolution of 0.5 mm/pix with a 35 mm lens (Nikon AF-S DX Nikkor f/1.8G). A Marumi filter polarizer was attached to the lens in order to attenuate reflective light spots. The fluid surface was illuminated using a high resolution projector (Epson Powerlite Home Cinema 8350) through another polarizer filter (Thorlabs). Images were treated using the prescription from Sepulvda-Soto~\cite{sepulveda2020} using Matlab. A typical outcome is shown in Fig. \ref{fig:setup} (right).

%
%

\begin{table*}[t]
\begin{ruledtabular}
 \begin{tabular}{lccccccc}
      {\bf Fluid}  & $\rho$ ($kg/m^3$)  & $T$ ($^{\circ} C$) & $\rho\nu$ ($Pa\times s$) & $K$ ($Pa \times s^n$) & n (-) & $\sigma$ ($N/m$)\footnote{Values taken from Hu~\cite{hu1991surface} and the Glycerine Producers’ Association~\cite{glycerine}.
}  \\[3pt] \hline  
       {\bf Water}   & 999 $\pm$ 3 & 18.5 $\pm$ 0.6 & 1.0e-3 $\pm$ 2e-04 & - & - & 0.074\\
       {\bf Glycerin}  & 1173 $\pm$ 1 & 18.4 $\pm$ 0.6 & 2.40e-2 $\pm$ 3e-04 & - & - & 0.075\\
       {\bf CMC}  & 1002 $\pm$ 1 & 18.1 $\pm$ 0.6 & - & 0.099 $\pm$ 0.004 & 0.76 $\pm$ 0.02 & 0.067\\
  \end{tabular}
  \end{ruledtabular}
  \caption{Physical properties of the used fluids. }
  \label{TableI}
\end{table*}

%
%
\section{Numerical solution}\label{sec:num}

In order to simulate numerically some aspects of the nonlinear wave propagation in the case of the CMC solution (which is non-newtonian as displayed in Fig.~\ref{fig:shear}), we used the the open-source CFD library OpenFOAM. We simulated 4 different non-Newtonian fluids, defined by theirs values of the fluid viscosity consistency $K$ and rheology non-linearity index $n$. To simplify and to speed-up the numerical procedure, a two-dimensional flow was considered (i.e., in the x-z plane). Thus, in the results we will present from these numerical simulations no transverse flow effects (y-direction following Fig.~\ref{fig:setup}) are taken into account, nor from the zero velocity condition applied on the lateral boundaries of the container or from its boundary layer. The solver OLAFOAM~\cite{higuera2015}, which is an upgrade from the numerical model IHFOAM, was used with a dynamic mesh scheme (olaDyMFoam). OLAFOAM is a numerical solver specially developed to solve wave dynamics at free interfaces between two fluid phases which supports the use of nonlinear viscosity coefficients (see Higuera's work~\cite{higuera2015} for more information of the numerical method).The fluid domain used was the same as the one used in experimental setup. The numerical mesh used in the simulations consists of 2 rectangular blocks, one for the working fluid on the bottom and another one for air on the top. Each block is constructed by a cubic mesh, with 0.69 mm discretization on the vertical direction and 0.4 mm discretization on the horizontal direction. At the boundaries of the numerical domain, mesh refinement is performed to better resolve boundary layers. Temporal discretization is fixed at 10$^{-4}$ s and the Courant number is set at 0.5.
The moving boundary condition at one of the vertical edges of the fluid domain, representing the moving wavemaker, was replaced by a time history of its position so as to generate equal amplitude solitary waves. On our runs, the fluid viscosity consistency $K$ and height to depth ratio $\varepsilon=H/h$ were kept constant while the rheology non-linearity index n ranged from 0.4 to 1.0. The OpenFOAM simulation type was set to “laminar”, neglecting the turbulent stresses in the Reynolds Averaged Navier-Stokes (RANS) equations. From these simulations, the wave profile is tracked for every mesh point in time, and thus the speed is computed by following the maximum of the wave train as it moves through the mesh.

\section{Results}\label{sec:res}

\subsection{Linear viscosity effects on wave celerity in non-linear long wave propagation}

\begin{figure*}
\centering
  \includegraphics[width=0.33\textwidth]{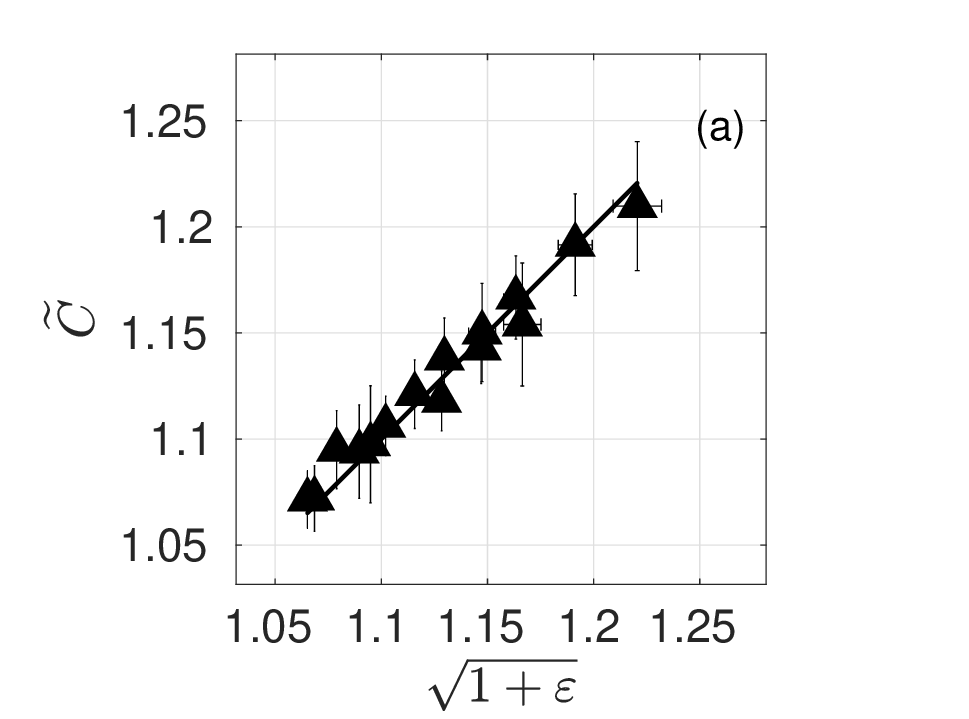}
   \includegraphics[width=0.33\textwidth]{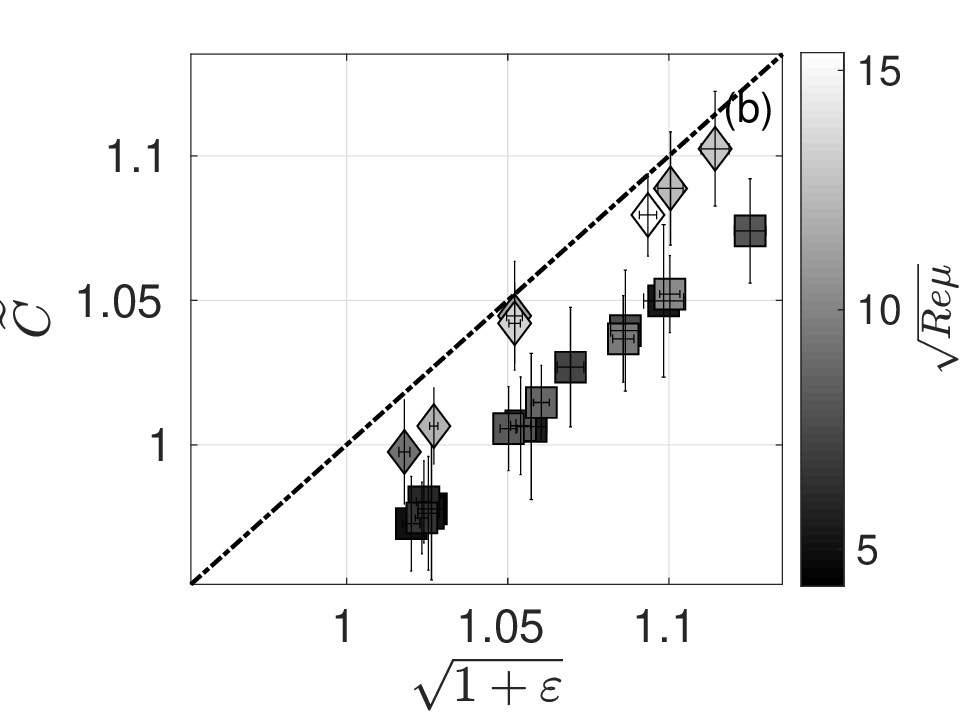}
 \includegraphics[width=0.33\textwidth]{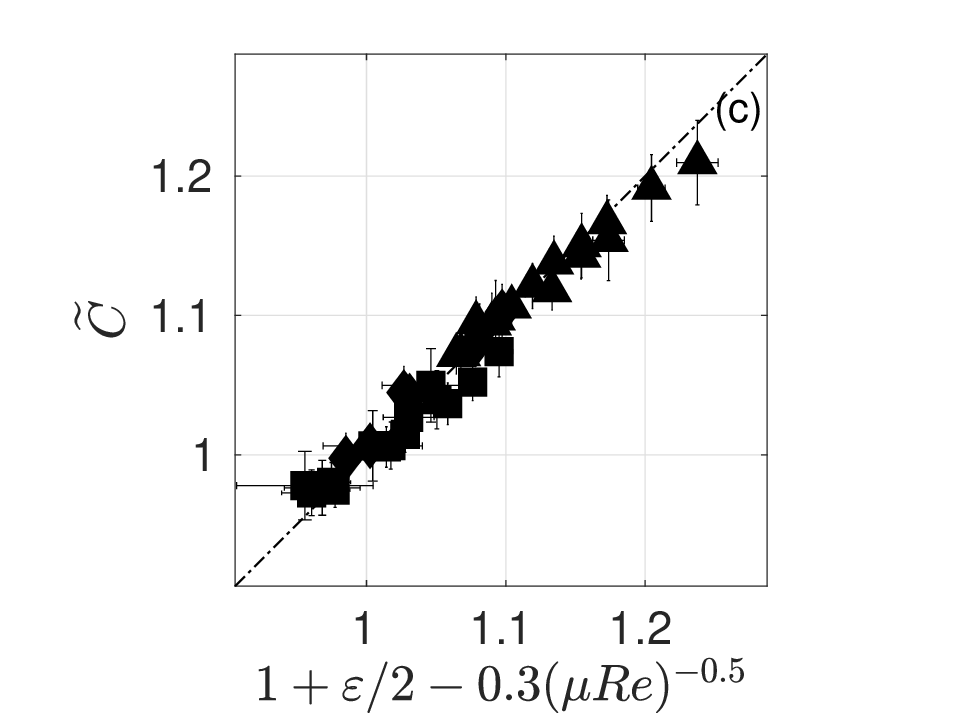}
 \caption{$(a)$ The measured dimensionless phase speed of solitary waves in water, $\tilde{C}$ ($\blacktriangle$) as a function of dimensionless wave elevation $\sqrt{1+\varepsilon}$ for $\varepsilon=0.14-0.49$ at still water depths $0.02-0.04$ (m). $(b)$ Comparison between experimental solitary wave phase speed in glycerin $0.7$ $\%$ $(\blacklozenge)$ and carboximetilcellulose $0.2$ $\%$ $(\blacksquare)$ aqueous solutions. Wave Reynolds number for Newtonian and non-Newtonian fluids  $\sqrt{\mu \bar{Re}}=(\mu \sqrt{gh} h /\nu)^{1/2}$, $\sqrt{\mu \bar{Re}}=\left(\rho \sqrt{gh} \mu h^n K^{-1} (\varepsilon \sqrt{gh})^{1-n} \right)^{\frac{1}{n+1}}$  (grayscale, left panel), respectively. Potential flow solitary wave speed line ($--$). $(c)$ Dimensionless wave phase speed in water ($\blacktriangle$), glycerin $0.7$ $\%$ $(\blacklozenge)$ and carboximetilcellulose $0.2$ $\%$ $(\blacksquare)$ aqueous solutions as a function of dimensional analysis dispersion relation including wave Reynolds number and one-to-one agreement (dashed line).}
\label{fig:celerity}
\end{figure*}

In order to characterize the linear viscosity effects in wave phase speed and attenuation in Newtonian fluids and the consistency index $K$ in non-Newtonian fluids, 45 laboratory experiments were carried out, and each one was replicated 5 times to estimate the variance of the results. Figure \ref{fig:celerity}$(a)$ shows the dimensionless wave phase speed $\tilde{C}$ as a function of the dimensionless wave height to depth ratio $(\varepsilon=H/h)$ for solitary waves in water (Newtonian). According to the Korteweg-de Vries solution~\citep{liu2004} the wave phase speed is equal to the square root of the dimensionless elevation ($\sqrt{1+\varepsilon}$). The non-linearity parameter $\varepsilon$ ranges from $0.14-0.49$ in order to maintain the weakly non-linear and dispersive validity of the Boussinesq theory~\cite{mei1998}. It is clear that the measurements agree with the inviscid Korteweg-de Vries solution very well. Thus, viscous effects are negligible. On the other hand, in figure \ref{fig:celerity}$(b)$ the measured wave phase speeds in glycerin (Newtonian) and carboximetilcellulose (non-Newtonian) fluids are plotted against the same dimensionless wave height in order to compare the higher linear and non-linear viscous effects on these wave celerity. The results are shown as a grayscale (left panel) of the Newtonian and non-Newtonian modified Reynolds numbers obtained from dimensional analysis~\cite{liu2006}. Comparing these results to the inviscid fluid wave celerity shows that wave speed decreases as the viscosity rises while keeping its nonlinearity constant. Figure \ref{fig:celerity}$(c)$ shows the dispersion relation obtained using dimensional analysis including the wave non-linear advective and viscous dissipative effects given by the equation \ref{eq:dispersion}, as described in~\ref{sec:dim_a}.  Although $\chi$ is a function of the non dimensional groups described above, in the range used in this study, it can be fitted as a constant $\chi=0.30\pm0.07$ in \ref{eq:dispersion} for all three fluids as a function of the modified Reynolds number. The resulting slopes using Eq. (\ref{eq:dispersion}) are $m \sim 0.999$ $\pm$ $0.002$ for water, $m=0.987$ $\pm$ $0.001$ for glycerin and $m=0.955$ $\pm$ $0.002$ for CMC at 2 $\%$. This relationship is consistent with higher Reynolds numbers or lower viscosities, where the viscous effects become negligible. Also, the modified Reynolds number for non-Newtonian fluids recovers exactly the Newtonian fluid Reynolds number for $n=1$ \citep{liu2004,johnson1997}. 


\begin{table*}[t]
\begin{ruledtabular}
 \begin{tabular}{lccccccc}
      {\bf Fluid}  & $h_{exp}$ ($m$) & $h_{num}$ ($m$) & $\varepsilon_{exp}$ & $\varepsilon_{num}$  & $c_{exp}$ $(m/s)$ & $c_{mum}$ $(m/s)$
  \\[3pt] \hline  
       {\bf Water}   & 0.040  & 0.040 & 0.207 & 0.208 & - & 0.074\\
       {\bf Glycerin}  & 1173 $\pm$ 1 & 18.4 $\pm$ 0.6 & 2.40e-2 $\pm$ 3e-04 & - & - & 0.075\\
       {\bf CMC}  & 1002 $\pm$ 1 & 18.1 $\pm$ 0.6 & - & 0.099 $\pm$ 0.004 & 0.76 $\pm$ 0.02 & 0.067\\
  \end{tabular}
  \end{ruledtabular}
  \caption{Experimental and numerical conditions associated with Fig.~\ref{fig:celerity}. }
\end{table*}

\subsection{Non-linear viscosity effects on wave celerity in non-linear long wave propagation}

\begin{figure*}
\centering
\includegraphics[width=0.33\textwidth]{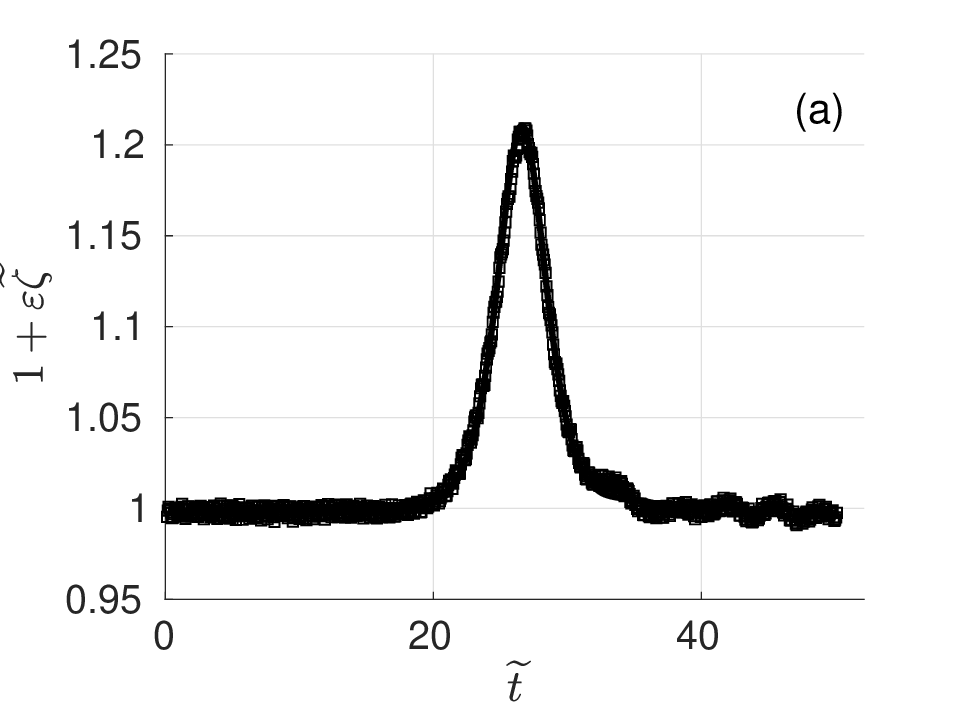}
\includegraphics[width=0.33\textwidth]{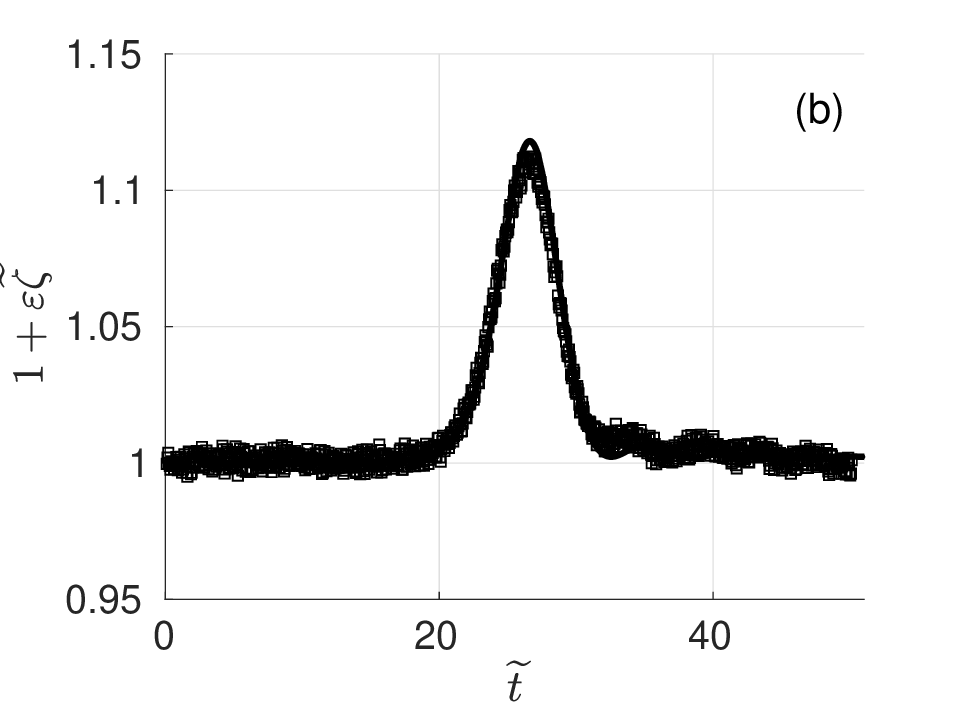}
\includegraphics[width=0.33\textwidth]{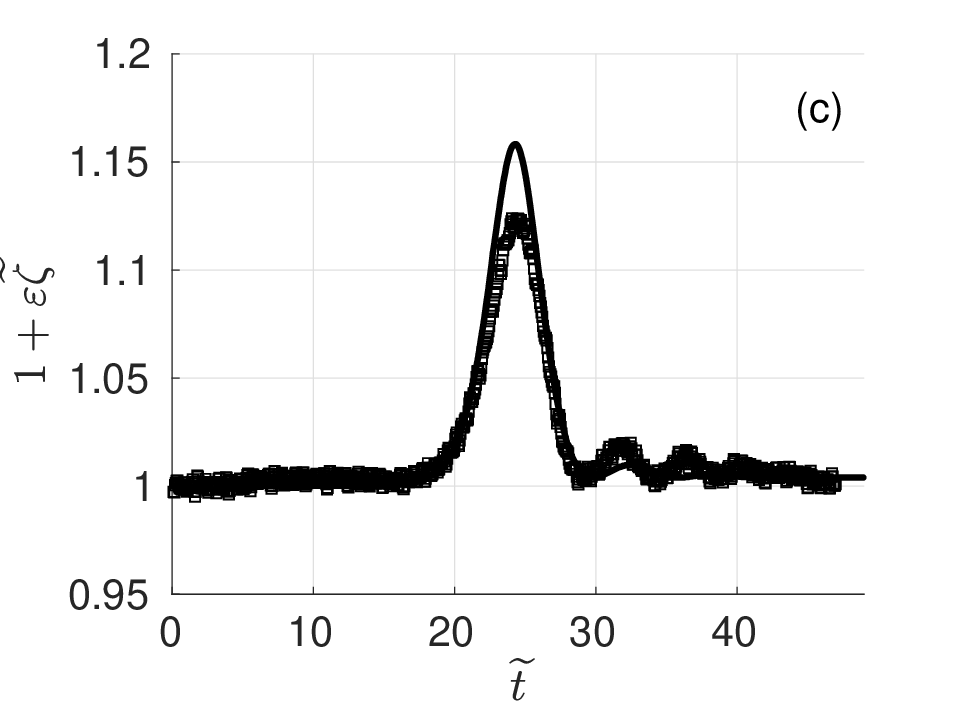}
\medskip
\includegraphics[width=0.33\textwidth]{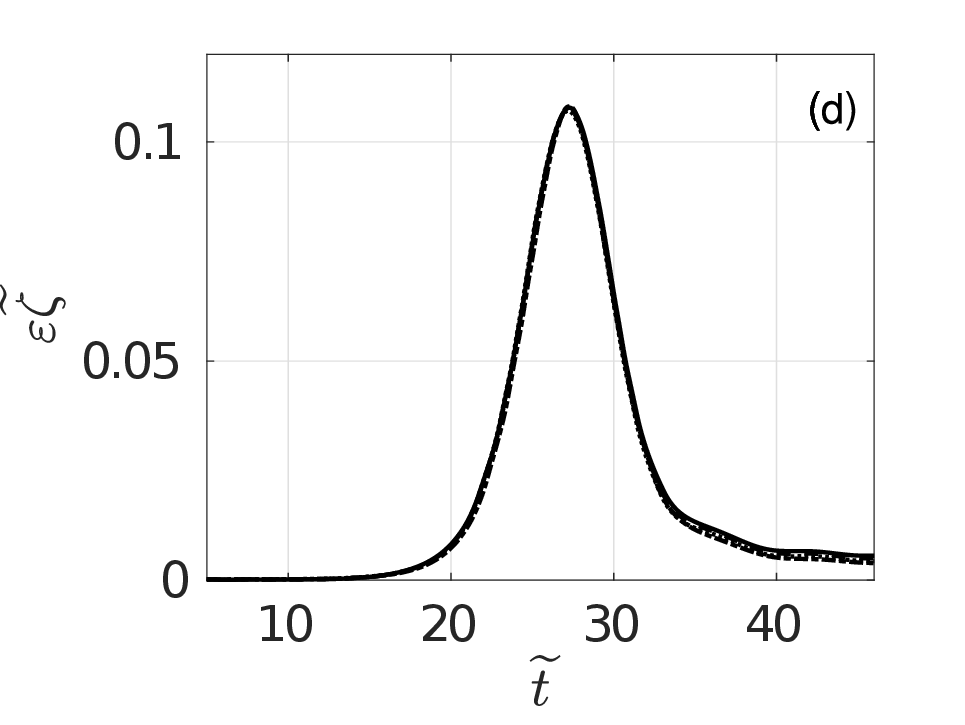}
\includegraphics[width=0.33\textwidth]{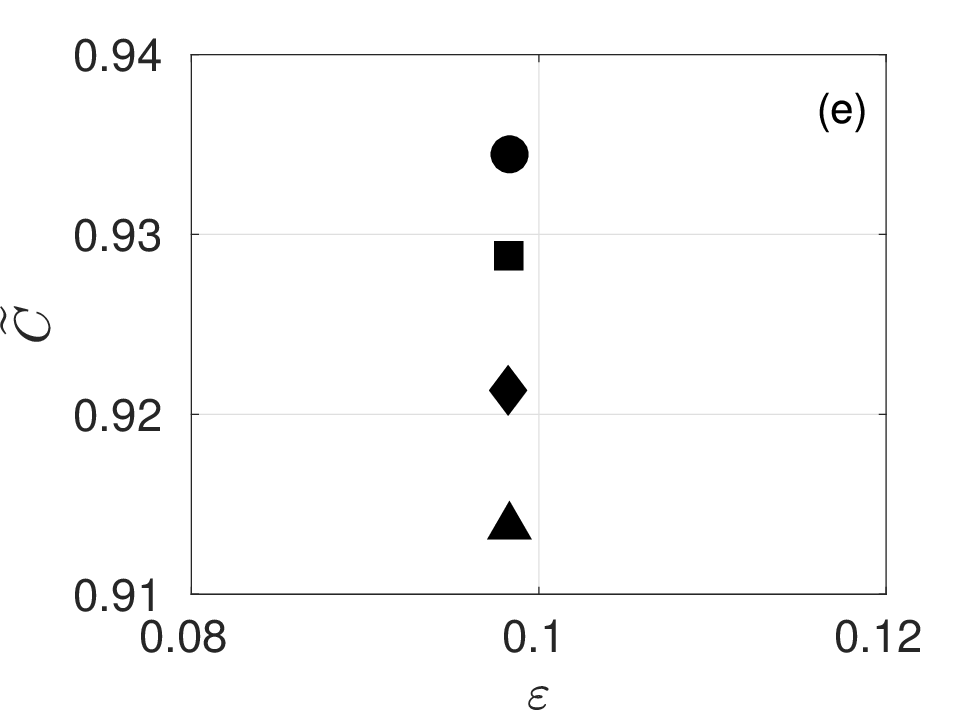}
\caption{ Experimental $(\blacksquare)$ and numerical $\mathbf{-}$ dimensionless free surface elevation wave profile $1+\varepsilon \tilde{\zeta}$ in dimensionless time $\tilde{t}=t \sqrt{gh^{-1}}$ for solitary waves in $(a)$ water, $(b)$ glycerin and $(c)$ carboxymethylcellulose solutions measured at 50 cm from the wave maker. $(d)$ Dimensionless free surface elevation wave profile $1+\varepsilon \tilde{\zeta}$ along dimensionless time $\tilde{t}$ at $x=0.6m$. Values of $\varepsilon=0.106$ and $h=0.04m$ for $n=0.4$ $(\bullet)$, $n=0.6$ $(\blacksquare)$, $n=0.8$ $(--)$ and $n=1$ $(-)$. $(e)$ dimensionless wave phase speed $\tilde{C}$ for constant $\varepsilon=H/h$ for different simulations at $n=0.4$ $(\bullet)$, $n=0.6$ $(\blacksquare)$, $n=0.8$ $(\blacklozenge)$ and $n=1$ $(\blacktriangle)$.}
\label{fig:nlcelerity}
\end{figure*}

To corroborate the non-linear viscosity effect (equation \eqref{eq:nn}) with the solitary wave phase speed observed experimentally, we carried out 4 different numerical simulations in which the non-linearity of the advective acceleration or wave height to depth $\varepsilon=0.10$ was kept constant as also the linear viscosity consistency $K=0.099$ $Pa s^n$ in equation \eqref{eq:nn}. Figure \ref{fig:nlcelerity}$(a),(b)$ shows the numerical solitary wave simulations for different rheologies at the same amplitude. Therefore, the influence of the non-linearity in fluid rheology on the wave phase speed was isolated as shown in the numerical results of figure \ref{fig:nlcelerity}$(b)$. The wave phase speed increases respect to the linear case $(n=1)$ at higher non-linearities $(n=0.4)$, where $n$ is the flow index in equation \eqref{eq:nn}. These results show a non-linear variation in wave phase speed along $n=0.4-1.0$ due to the high non-linearity of the mass and momentum conservation described in equations \ref{eq:continuity} and \ref{eq:momentum}. Accordingly, as we show from figure \ref{fig:nlcelerity}$(a)$ the solitary wave wave profile is preserved in different rheological media, keeping the wave height and viscosity consistency constant in order to quantify the varying rheology effect on wave phase speed. In this same matter, realistic flow index values of $(n)$ in the range of $0.4-1.0$ \cite{dodge1959} were used to simulate different fluid rheologies. Additionally, three validation cases with experimental data were employed for water, glycerin and carboxymetilcellulose experiments yielding an error of $1\%$, $4\%$ and $5\%$ in wave phase speed, and $1\%$, $11\%$ and $19\%$ in wave height, respectively. In this context, the phase speed error is $\mathcal{O}(\Delta \varepsilon/2)$ and the wave height differences in pseudoplastic fluids may due to the bidimensionality simplification used for the numerical model described in \ref{sec:num}, which did not include the sidewalls boundary layer influence. Non-slip boundary conditions and high viscosity in a confined transverse direction will decrease the wave height in a considerable manner, which is observed experimentally. This is not modeled in the bi-dimensional simplification of our numerical simulations. Further work on the boundary layer simulation is needed, but remains out of the scope of this work.

\subsection{Non-linear viscosity effects on wave damping in non-linear long wave propagation}

\begin{figure*}
\centering
  \includegraphics[width=0.45\textwidth]{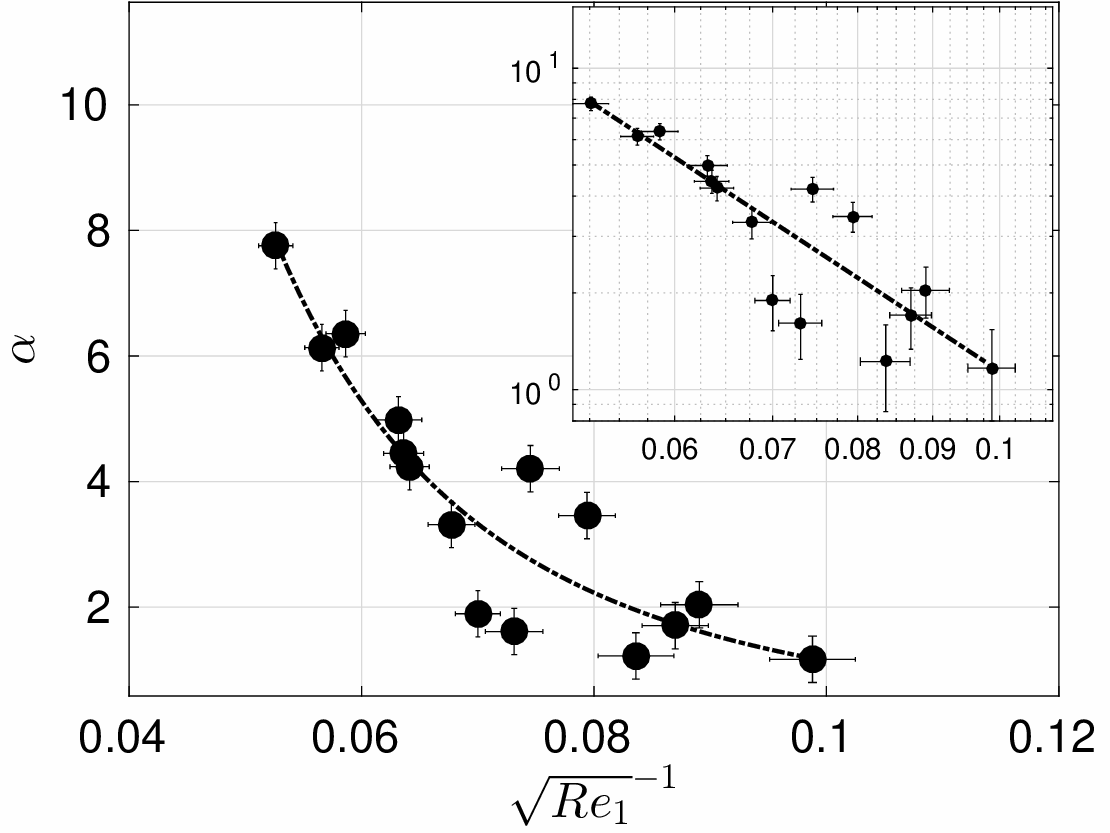}
  \includegraphics[width=0.45\textwidth]{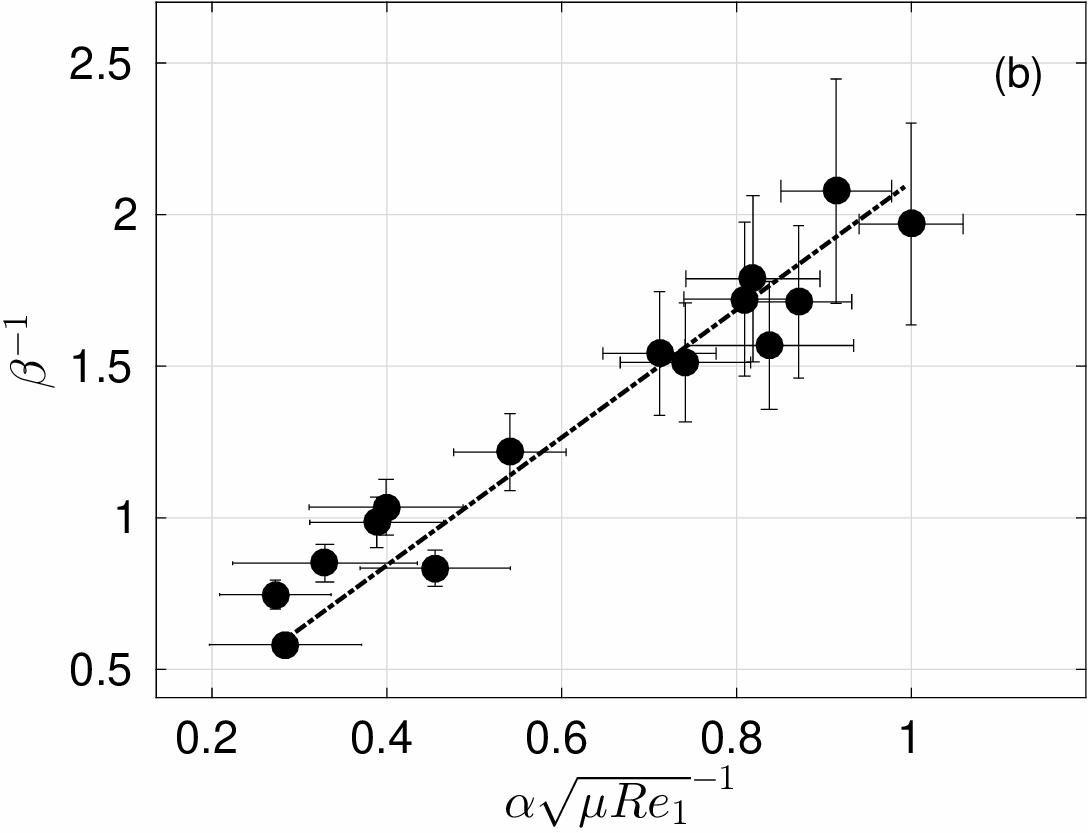}
  \caption{Results of experimental non-linear wave damping in non-Newtonian fluids. $(a)$ Wave attenuation $\alpha$ as a function of the inverse square root of $\bar{Re}_1$.  $(b)$ Exponent dependence with $\bar{Re}_1=\mu^{\frac{n-1}{2}}\bar{Re}^{\frac{n+1}{2}}$. Non-linear regressions from experimental data yields $\alpha=0.001 {\bar{Re}_1}^{3/2}$ and $\beta^{-1}=2.108 \alpha (\mu \bar{Re}_1)^{-1/2}$.}
\label{fig:damping}
\end{figure*}

Completing the characterization of solitary wave propagation in non-Newtonian fluids, we measured the wave damping coefficients as a function of viscosity, referred to the theoretical relation for the dimensionless free surface $\zeta/H=(1+\alpha t)^{-\beta}$ obtained for weakly viscous Newtonian fluids by Liu~\cite{liu2004}. In figures \ref{fig:damping}$(a)$ and \ref{fig:damping}$(b)$ experimental dimensionless numbers relations for the wave attenuation coefficients for the damping coefficients are shown for non-linear waves in non-Newtonian fluids. Experimental data for wave damping was fitted by non-linear least squares regression obtaining the relations $\alpha=0.001 {\bar{Re}}^{3/2}$ and $\beta^{-1}=2.108 \alpha (\mu \bar{Re}_1)^{-1/2}$, where $\bar{Re}_1=\mu^{\frac{n-1}{2}}\bar{Re}^{\frac{n+1}{2}}$ is a scaled Reynolds number appearing directly in the calculation of Liu~\cite{liu2004}. From the expression above, we can see that $\bar{Re}_1$ depends on $\bar{Re}$, $\mu$ and $n$. Finally, using the above results, it yields a dependence between the damping power, $\bar{Re}$ and $\mu$ in the form of $\beta^{-1} \sim 0.002 \bar{Re}_1 / \mu$. 
%

\section{Discussion}\label{sec:dis}

The experimental and numerical results presented in this paper provide a quantitative description and understanding of the linear and non-linear viscous effects on the wave phase speed and attenuation for Newtonian and non-Newtonian fluids. In this context, we addressed the non-linear wave phenomena using an approach based on the complete Navier-Stokes equations to analyze viscous effects on the non-linear wave dispersion relation, overcoming the limitations of existent studies~\cite{connolly2015}. We also summarize the viscous effects on the wave phase speed as a dispersion relation (equation \ref{eq:dispersion}) which is valid for different fluid rheologies at laminar flow regime in the accessible range of wavelengths as shown in figure \ref{fig:setup}$(c)$.  

In the context of linear viscous effects on non-linear wave phase speed, the resulting slope of unity (figure \ref{fig:celerity}$(a)$) for the dimensionless phase speed as a function of the dimensionless wave elevation in the water solitary wave is in good agreement with the wave celerity obtained for the potential flow solution from the Korteweg-de Vries equations. The latter  shows that the wave phase speed of solitary waves in water is weakly affected by the viscous effects and is mainly driven by the advective wave non-linearity rather than viscosity in this case. In contrast, the lower slopes for glycerin and CMC at $\%$ (figure \ref{fig:celerity}$(b)$) obtained for the relation of the dimensionless wave celerity and wave elevation are shifted downward showing a progressive increase in the viscosity (at a fixed shear rate larger than 1 s$^{-1}$) reduces the wave phase speed, decelerating the fluid velocity due to the presence of higher viscous stress forces. Furthermore, at a shear rate of unity the dynamic viscosities of the glycerol and the non-Newtonian fluids are 24 $Pa\; s$ and 99 $Pa\; s$, respectively. As a result of this, according to the dispersion relation obtained (equation \ref{eq:dispersion}), our analysis suggests that a higher viscosity lowers the Reynolds number, resulting in a more laminar flow regime, which decreases the wave phase speed by $\mathcal{O}(\bar{Re}^{-0.4})$. The latter is consistent with the linear dispersion relation for linear waves in viscous fluids \citep{johnson1997}. In contrast, at lower viscosities and higher Reynolds numbers as is the case of water, the viscous contribution in equation \ref{eq:dispersion} becomes negligible, result consistent with the data presented in figures \ref{fig:damping}$(a)$ and \ref{fig:damping}$(c)$ for water. It must be noticed that we have used shear rates of order unity as a lower limit.  Experimentally, $\dot{\gamma}$ in our configuration are estimated to be in the range between 10-40 s$^{-1}$. We estimate $\dot{\gamma}\sim c(h+H))$, using the experimentally measured values presented above. Although this estimation gives us a bulk value of $\dot{\gamma}$ and it can be considered a crude one, it gives us an idea of the validity of using the power law model for shear rates of that order and larger. The above estimation thus presents a lower boundary of the values of $\dot{\gamma}$. Of course, this analysis is valid once the wave is arrived and the fluid has been set in motion, which is confirmed by our numerical simulations. 

The relation (equation \ref{eq:dispersion}) for the five dimensionless celerity $(\widetilde{C})$, Reynolds ($\bar{Re}$), shallowness parameter ($\mu$), height to depth ratio ($\varepsilon$) and flow index $(n)$  was validated experimentally for a wide range of realizations, comparing the measured wave phase speed, non-linear terms and Reynolds number form the equation \ref{eq:dispersion} in figure \ref{fig:celerity}$(c)$. This result holds for both Newtonian and non-Newtonian fluids. In fact ,imposing the flow index as linear viscosity for a Newtonian fluids $(n=1)$, equation \ref{eq:dispersion} retrieves the Reynolds number for Newtonian fluids derived by Liu \cite{liu2004} and Mei~\cite{mei1998}, among others. Moreover, imposing the linear limit for linear waves, the linear dispersion relation is recovered \citep{johnson1997}.  

Regarding the viscous non-linear effects on the wave speed, our numerical simulations shows that lower power indexes in the non-linear viscosity of the fluid shear stress (equation \ref{eq:nn}) or in the same way, higher non-linearities, increases the wave phase speed (figure \ref{fig:celerity}$(b)$) due to the shear-thinning nature of the fluid rheology, which at higher flow indexes ($n$) the effective fluid viscosity decreases, reducing its resistant and dissipative shear forces, thus accelerating fluid motion. It is observed that this slowdown in the wave phase speed is non-linear due to the non-linearity of the Navier-Stokes equations (equation \ref{eq:momentum}). Moreover, the general relation \ref{eq:dispersion} predicts that the wave phase speed in a shear thickening or dilatant non-Newtonian fluid $(n>1)$ will decelerate at higher flow index values $(n)$, as a result from its higher effective non-linear viscosity. Inability to include the sidewalls boundary layers may induce numerical innacuracies in higher viscous fluid simulations.

Finally, according to the wave damping relation obtained through dimensional analysis we find that the damping coefficient decreases at lower viscosities (higher Reynolds numbers) and increases for less shallow waves, which is consistent with the theoretical relation for low viscosity Newtonian fluids~\cite{liu2004,liu2006}. Thus, from our results, a possible rheometric technique can be conceived where nonlinear wave propagation on non-Newtonian fluids can be used to probe and measured fluid properties such as $K$ and $n$ which can be compared and contrasted with other standard rheometric instruments and techniques. 

\section*{Data Availability Statement}
Data is available upon request to C. F. and A. T. Details of the deduction of the equations presented in the paper can be found in the Master's Thesis of C. C.~\cite{tesisCC}\\

\begin{acknowledgments}
C. C. acknowledges the financial support given by the scholarship CONICYT-PCHA/MagísterNacional/2016 22161261 (Chile) and the funding from the fellowship International Engagement Graduate Department of the University of Chile. A. T.acknowledges the funding from FONDECYT grant 1161751 (Chile) and ANID Project AFB230001.  Powered@NLHPC: This  research  was  supported  by the supercomputing infrastructure of the National Laboratory for High Performance Computing (NLHPC) (ECM-02) (Chile).
\end{acknowledgments}

\end{document}